\documentclass[journal]{IEEEtran}
\usepackage{amsmath,amsfonts}
\usepackage{algorithmic}
\usepackage{array}
\usepackage[caption=false,font=normalsize,labelfont=sf,textfont=sf]{subfig}
\usepackage{textcomp}
\usepackage{stfloats}
\usepackage{url}
\usepackage{verbatim}
\usepackage{graphicx}
\usepackage{bm}
\usepackage{multirow}
\usepackage{makecell}
\usepackage{booktabs}
\usepackage{color}
\usepackage{adjustbox}
\usepackage{cite}

\usepackage{balance}
\begin{document}
\bstctlcite{settingbib}

\title{LLM4WM: Adapting LLM for Wireless Multi-Tasking}
\author{\IEEEauthorblockN{
		Xuanyu Liu,~Shijian Gao,~\IEEEmembership{Member,~IEEE,}~Boxun Liu,~\IEEEmembership{Graduate Student Member,~IEEE,}
        \\Xiang Cheng,~\IEEEmembership{Fellow,~IEEE,}~Liuqing Yang,~\IEEEmembership{Fellow,~IEEE}
        }
\thanks{
Xuanyu Liu, Boxun Liu and Xiang Cheng are with the State Key Laboratory of Advanced Optical Communication Systems and Networks, School of Electronics, Peking University, Beijing 100871, China (e-mail: xvanyvliu@gmail.com; boxunliu@stu.pku.edu.cn; xiangcheng@pku.edu.cn).
			
Shijian Gao is with the Internet of Things Thrust, The Hong Kong University of Science and Technology (Guangzhou), Guangzhou 511400, China (e-mail: shijiangao@hkust-gz.edu.cn).
		
Liuqing Yang is with the Internet of Things Thrust and Intelligent Transportation Thrust, The Hong Kong University of Science and Technology (Guangzhou), Guangzhou 511400, China, and also with the Department of Electronic and Computer Engineering and the Department of Civil and Environmental Engineering, The Hong Kong University of Science and Technology, Hong Kong, SAR, China (e-mail: lqyang@ust.hk).
}} 

\maketitle

\begin{abstract}
The wireless channel is fundamental to communication, encompassing numerous tasks collectively referred to as channel-associated tasks. These tasks can leverage joint learning based on channel characteristics to share representations and enhance system design. To capitalize on this advantage, LLM4WM is proposed—a large language model (LLM) multi-task fine-tuning framework specifically tailored for channel-associated tasks. This framework utilizes a Mixture of Experts with Low-Rank Adaptation (MoE-LoRA) approach for multi-task fine-tuning, enabling the transfer of the pre-trained LLM's general knowledge to these tasks. Given the unique characteristics of wireless channel data, preprocessing modules, adapter modules, and multi-task output layers are designed to align the channel data with the LLM's semantic feature space. Experiments on a channel-associated multi-task dataset demonstrate that LLM4WM outperforms existing methodologies in both full-sample and few-shot evaluations, owing to its robust multi-task joint modeling and transfer learning capabilities.
\end{abstract}

\begin{IEEEkeywords}
large language models, Mixture of Experts, Low-Rank Adaptation, multi-task learning, wireless multi-tasking, transfer learning.
\end{IEEEkeywords}

\section{Introduction}
\IEEEPARstart{T}{he} quality and reliability of communication are largely determined by the wireless channel, which plays a crucial role in this process. To achieve low latency and high reliability, millimeter-wave (mmWave) and Multiple-Input Multiple-Output (MIMO) technologies are among the most promising solutions \cite{larsson2014massive, lu2014overview, rusek2012scaling}. Maximizing the benefits of these technologies relies heavily on accurate channel estimation \cite{gao2020estimating}. However, this task becomes increasingly challenging as the number of antennas grows and environmental dynamics become more complex. Understanding the wireless channel's characteristics—such as fading, interference, and multipath propagation—is crucial for optimizing communication performance, and also supports other related technologies, such as the design of integrated sensing and communications (ISAC) systems \cite{liu2024beam}. Fortunately, artificial intelligence (AI) has significantly improved channel estimation accuracy, fostering optimism for the practical implementation of advanced wireless applications \cite{soltani2019deep, balevi2020high, arvinte2022mimo}. AI also enhances various communication tasks, including channel prediction, beamforming, and positioning, demonstrating notable effectiveness and robustness.

\begin{figure*}[htbp]
    \centering
    \includegraphics[width=1\linewidth]{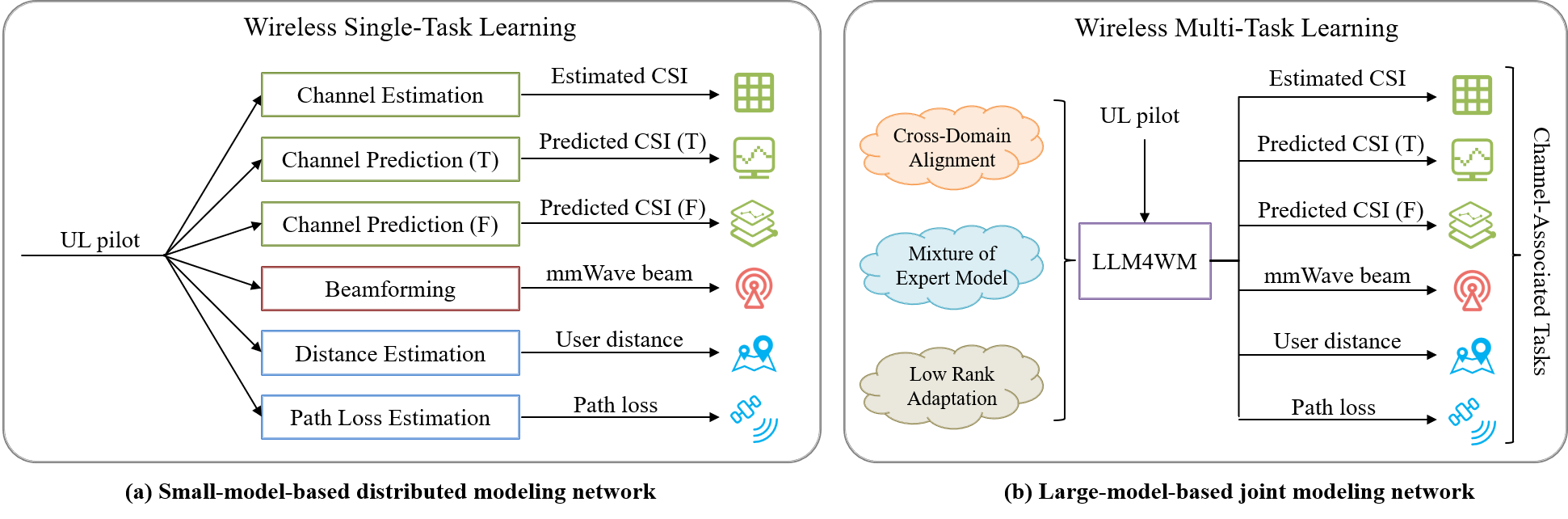}
    \caption{An illustration highlighting the differences in the workflows between a small-model-based distributed modeling and a large-model-based joint modeling.}
    \label{workflow}
    \end{figure*} 

Although AI has demonstrated significant potential in communication systems, existing AI-powered communication methods still encounter several issues. 
First, AI approaches often require large amounts of high-quality data, and collecting such datasets can impose substantial communication overhead on the system. Additionally, due to generalization issues, AI models need to be retrained in response to dynamic changes in the environment, further increasing the communication burden. Furthermore, existing AI methods frequently struggle in complex and highly dynamic scenarios, often due to the limited scale of the models.

To address these challenges, \cite{cheng2023intelligent} proposed that multi-modal sensing could effectively capture the propagation characteristics of the wireless channel, a concept summarized as the Synesthesia of Machines (SoM). This approach aims to optimize communication systems by leveraging multi-modal sensing to enhance their design and performance \cite{zhang2024integrated, yang2024synesthesia}.

Another potential optimization approach is the introduction of multi-task learning. Given that the wireless channel is a critical component of the wireless communication process, and many communication tasks focus on the extraction and utilization of channel features under various conditions, jointly learning these channel-associated tasks can yield significant training benefits by extracting shared channel representations across tasks. For instance, in \cite{jagannath2021multi}, two key tasks of wireless signal recognition—signal classification and modulation recognition—were jointly trained. In \cite{xie2021multi}, the joint training and inference of direct channel and cascaded channel estimation in reconfigurable intelligent surface systems were proposed, effectively reducing pilot overhead. Despite their effectiveness, these methods have limitations, such as issues with data imbalance and the seesaw effect in multi-task learning approaches that rely on shared representations at the bottom layers. Additionally, challenges arise in scaling the number and diversity of tasks due to limited model capacity, as most methods combine only two closely related tasks.

Recently, Large language models (LLMs) have become powerful learners adept at handling multiple tasks,
demonstrating remarkable reasoning capabilities and generalization across various domains, including natural language processing (NLP) \cite{brown2020language}, healthcare \cite{singhal2023towards}, law \cite{colombo2024saullm}, and finance \cite{wu2023bloomberggpt}. The GPT-4 has shown outstanding performance in NLP tasks,
while TTM \cite{ekambaram2024tiny} has also demonstrated remarkable few-shot and zero-shot learning abilities in time series processing tasks. Inspired by these successes, there is growing interest in leveraging pre-trained models for cross-domain tasks, including channel-associated tasks in wireless communication. For instance, an LLM-empowered channel prediction method LLM4CP was proposed in \cite{liu2024llm4cp}, achieving greatly improved few-shot generalization ability. A foundational channel model WiFo \cite{liu2024wifo} was trained on a diverse channel dataset to perform tasks like time-domain and frequency-domain prediction with zero-shot learning, but they primarily focus on channel reconstruction. Our objective is to enhance multiple channel-associated tasks in wireless communication using LLMs, which poses challenges such as effectively transferring cross-domain knowledge and managing task diversity. We address these challenges through the MoE-LoRA fine-tuning method \cite{liu2024moe}, designed for multi-task scenarios, while noting that recent work has not sufficiently considered the specific relationships between channel-associated tasks,
which may limit the model's capacity to acquire general representations.

Specifically, we propose a wireless multi-task fine-tuning framework leveraging pre-trained large language models, which is successfully applied to the channel-associated multi-tasks. Unlike previous multi-task learning approaches that rely on shared bottom layers, we freeze most of the large model’s parameters and introduce MoE-LoRA for wireless multi-task fine-tuning. On the one hand, tasks can share experts' weight, which helps the network learn common knowledge across tasks, while on the other hand, the independence of experts and the gating mechanism ensure the differentiation of task-specific features. Additionally, we incorporate multi-task adapter modules at both the input and output layers of the LLM to ensure alignment between the feature space of communication tasks and the semantic space of the pre-trained LLM. We also design a pre-processing method and corresponding output header for each task to enhance our framework further, optimizing the model's overall performance and adaptability. The core contributions of this paper are summarized below:

\begin{itemize}
\item LLM4WM is introduced as a novel method that leverages LLM to facilitate wireless multi-tasking. This approach represents a pioneering effort in fine-tuning LLM using MoE-LoRA to extract a joint representation tailored for wireless multi-task scenarios, establishing a new standard in this area of research.
\item To ensure effective cross-domain alignment, a customized pre-processing method and corresponding output header have been developed for each task. Additionally, multi-task adapters are created to bridge the gap between the LLM’s semantic feature space and the specific feature space of wireless tasks, enhancing adaptability and performance. 
\item The model exhibits excellent performance on a range of wireless communication tasks, including channel estimation, channel prediction, localization enhancement, and beam management. Furthermore, it exhibits impressive generalization capabilities, highlighting robustness and versatility for diverse applications in the wireless domain.

\end{itemize}
   
\textit{Notation:} $(\cdot)^{\rm H}$, $\lvert \cdot \rvert$ and $\Vert\cdot\Vert$ denote the conjugate transpose, determinant and $l_2$ norm, respectively. $\bm{a}[i]$ is the $i\mbox{-}$th element of a vector $\bm{a}$ and $\bm{M}[i,j]$ denotes the element of matrix or tensor $\bm{M}$ at the $i\mbox{-}$th row and the $j\mbox{-}$th column.
The slicing operation $\bm{M}[s_1 : e_1 : i_1, s_2 : e_2 : i_2]$ is used to extract a submatrix or sub-tensor from \(\bm{M} \). Here, $s_1$ and $s_2 $ represent the starting indices, $e_1$ and $e_2$ denote the ending indices, and $i_1$ and $i_2$ are the step sizes that determine the interval between selected indices. If either $i_1$ or $i_2$ equals 1, the step size can be omitted for simplicity.
$\mathbb{E}\{\cdot\}$ denotes the statistical expectation of the enclosed variable or expression.

\section{SYSTEM DESCRIPTION}

A dual-frequency communication system operating at sub-6G and mmWave frequencies is considered, which consists of one base station (BS) and one user equipment (UE). 
Both the BS and UE are equipped with transceivers for both frequency bands, with multiple-input multiple-output (MISO) and orthogonal frequency division multiplexing (OFDM) technologies applied to each. The sub-6G and mmWave antennas are co-located and aligned, with similar apertures, enabling them to share spatial features. 
For clarity, we use $\tilde{\left( \cdot \right)}$ notation, such as $\tilde{x}$ to denote parameters related to the sub-6G system. In mmWave band, the BS adopts an analog beamforming architecture and features $N_t$ antennas arranged in a uniform linear array (ULA) configuration, 
while in the sub-6G band, it utilizes a fully digital beamforming architecture with \( \tilde{N}_t \) ULA antennas. The UE is equipped with an omnidirectional antenna for both frequency links and can accommodate multiple antennas through parallel processing. 
 
\subsection{Channel Model}
For both sub-6G and mmWave channels, the classical cluster-based multipath channel model is utilized to describe the downlink and uplink CSI between the BS and user at time $t$ and frequency $f$:
\begin{equation}
\resizebox{0.9\hsize}{!}{$
\begin{aligned}\label{CSI}  
&\bm{h}(t,f)=\sum_{n=1}^N \sum_{p=1}^{P_n}\beta_{n,p}e^{j[2\pi (\upsilon_{n,p} t-f\tau_{n,p})+\Phi_{n,p}]}\bm{a}(\theta_{n,p}).
\end{aligned}$}
\end{equation}
In this context, $N$ and $P_n$ are the number of clusters and paths in each cluster, respectively.
$\beta_{n,p}$, $\upsilon_{n,p}$, $\tau_{n,p}$, and $\Phi_{n,p}$ represent the complex path gain, doppler frequency shift, delay, and random phase, respectively.
$\bm{a}(\theta_{n,p})$ represents the steering vector of the corresponding path, where $\theta_{n,p}$ denote the azimuth angles of departure (AoD).
Considering the structural characteristics of ULA, the expression for $\bm{a}(\theta_{n,p})$ is derived as:
\begin{equation}
\resizebox{0.9\hsize}{!}{$
\begin{aligned}\label{aodvector}  
\bm{a}(\theta_{n,p})=[1,e^{j \frac{2\pi fd_{\rm v}\sin (\theta_{n,p})}{c}},...,e^{j\frac{2\pi (N_{\rm t}-1) fd_{\rm v}\sin (\theta_{n,p})}{c}}] .
\end{aligned}$}
\end{equation}
Here, $d_{\rm v}$ represents the antenna spacing in the vertical direction.

\subsection{Signal Model}
For the mmWave links, We consider a downlink MISO-OFDM signal transmission process, where $K$ subcarriers are activated, with the $k\mbox{-}$th subcarrier denoted as $f_k$.
According to Eq. (\ref{CSI}), the downlink CSI at time $t$ and the $k\mbox{-}$th subcarrier is $\bm{h}_{t, k}=\bm{h}(t,f_k)$, which can be obtained through channel estimation or prediction.
Considering transmit precoding at the BS side, the received downlink mmWave signal of the time t and $k\mbox{-}$th subcarrier at the user side is derived as:
\begin{align}
\bm{y}_{t,k}=\bm{h}_{t, k}^{\rm H}\bm{w}_{t} {x}_{t, k} + \bm{n}_{t, k},
\end{align}
where $\bm{w}_{t} \in \mathbb{R}^{{N}_{v}\times1}$ is the beam vector which is selected from predefined codebooks ${\bm{\mathcal{W}}}$, i.e. $\bm{w}_{t}\in{\bm{\mathcal{W}}}$. And $\bm{n}_{t, k}$ is the additive white gaussian noise (AWGN).
The achievable spectral efficiency (SE) \cite{marzetta2016fundamentals} of the downlink transmission process is derived as:
\begin{align}\label{SE}
R_t=\sum_{k=1}^{K_{\rm s}}\log_2{\left( 1+\frac{\lvert\bm{h}_{t,k}^{\rm H}\bm{w}_{t}\rvert^2}{\sigma_n^2} \right)}.
\end{align}

Through beam training, all beam vectors are traversed from the codebook, and the one with the highest SE is selected as the optimal beam vector, which can be formulated as:
\begin{align}\label{Angleprecoding}
  \bm{w}_{t}^*=\arg\max_{\bm{w}_{t}\in\bm{\mathcal{W}}} R_t. 
\end{align} 

Similarly, the sub-6G downlink signal transmission process also adopts MISO-OFDM, and it has the same transmission expression as the mmWave link, which will be omitted here for brevity. However, the sub-6G link employs digital precoding, so 
the uplink channel at the pilot positions is estimated using the Least Squares (LS) method as follows:
\begin{align}
   \tilde{\bm{h}}_{LS}(t, f_k) =  \bm{\tilde{y}}_{t, k} / \tilde{x}_{t, k},  
\end{align}
where $\tilde{x}_{t, k}$ and $\tilde{y}_{t, k}$ represent the uplink pilot signal sent by the user and the signal received by the base station respectively. Then to maximize the system SE, the matched filtering based precoding is applied as follows:
\begin{align}\label{mf}
\bm{\tilde{w}}_{t, k}^*= \frac{\bm{\tilde{h}}_{LS}(t, f_k)}{ \Vert\bm{\tilde{h}}_{LS}(t, f_k)\Vert} ,
\end{align}
where $\tilde{w}_{t, k}^*$ represents the beam vectors of time $t$ and the $k\mbox{-}$th subcarrier at sub-6G link. Notably, the effectiveness of $\tilde{w}_{t, k}^*$ depends on the accuracy of $\tilde{{h}}_{LS}(t, f_k)$.
Specifically, inaccurate $\tilde{{h}}_{LS}(t, f_k)$ will lead to mismatched $\tilde{w}_{t, k}^*$ with the actual $\tilde{h}(t, f_k)$, reducing the received signal-to-noise ratio (SNR) and thereby impairing the SE.
Therefore, accurate CSI is vital for improving the SE of the system.

\section{TASK DESCRIPTION}
To achieve higher SE in the sub-6G frequency band, accurate channel matrix estimation from pilot signals is essential \cite{soltani2019deep}. Research has focused on enhancing pilot-based channel estimation for massive antenna arrays and addressing channel aging in high mobility scenarios through time-domain prediction \cite{jiang2022accurate}. Additionally, some studies reduce pilot overhead by extrapolating in the frequency domain \cite{safari2019deep}.
In the mmWave frequency band, the increased number of antennas necessitates improved beam scanning efficiency. Researchers utilize the spatial correlation between frequency bands to estimate optimal mmWave beams based on the sub-6G channel matrix \cite{lv2024sub}. Parameters like user distance $x_d$ and path loss $x_{pl}$ are also critical for configuring communication links, with some works estimating these factors from channel data \cite{salihu2022attention, sun2022environment}. Although the objectives and optimization methods of these tasks differ, they all leverage characteristics of sub-6 GHz channels. Moreover, base stations often need to perform these tasks simultaneously for efficient communication system operation. Thus, designing a unified network to address these tasks concurrently is essential.

We categorize these tasks into three classes: channel reconstruction, beam management, and radio environment mining. To enhance clarity and understanding, we define \(\mathcal{T} = \{ {CE}, {CP}, {PF}, {BF}, {DE}, {PE} \} \) as the task set. The detailed descriptions are presented in Tab. \ref{task description}. 
\begin{table}[h]
\renewcommand\arraystretch{1.5}  
\caption{Description and classification of task id}
\label{task description}
\centering
\scriptsize
\begin{tabular}{c|c|c}
\hline
\textbf{Task class} & \textbf{Task id} & \textbf{Task content} \\ 
\hline
\multirow{3}{*}{ \makecell[c]{Channel \\ Reconstruction }} 
& CE & Channel estimation \cr
& CP & Temporal domain channel prediction \cr
& PF & Frequency domain channel prediction \cr
\hline
\multirow{2}{*}{ \makecell[c]{Beam \\ Management }} & \multirow{2}{*}{BF} & \multirow{2}{*}{Sub-6G assisted mmWave beamforming} \cr
& & \cr
\hline
\multirow{2}{*}{ \makecell[c]{ Radio \\ Environment Mining }} 
& DE & Distance estimation \cr
& PE & Path loss estimation \cr

\hline
 
\end{tabular}
\end{table}
 
Each task corresponds to a dataset \( D_n = \{ X^{I}_n, X^{L}_n \} \), where \( X^{I}_n \) represents the input samples for the task \( n \)  and \( X^{L}_n \) represents the labels for the task \( n \). All tasks can be formulated in the following form:
\begin{subequations}\label{problem}
\begin{align}
\max_{{\Omega}_n}\quad & {\text{Score}_n}=\mathbb{E}\left\{ f_{eval_n}(\bm{X}^{L}_n, \bm{X}^{O}_n)  \right\} \label{NMSE}\\
s.t.\quad& \bm{X}^{O}_n=f_{{\Omega}_n}( \bm{X}^{I}_n ), \quad n \in \mathcal{T},
\end{align}
\end{subequations}
where $f_{eval_n}$ and $f_{{\Omega}_n}$ are the evaluation function and the constructed mapping function for the task \( n \), respectively. $\bm{X}^{O}_n$ represent the output results of the model for the task \( n \). Fig. \ref{workflow} illustrates the complete workflow of the system. Detailed descriptions of the specific sub-tasks will follow.

\subsection{Channel Reconstruction: Interpolation and Prediction} 
Channel reconstruction tasks typically aim to use known channel matrices to predict or interpolate the target channel matrices. The model is expected to capture the inter-domain correlations of the channel matrix across time, frequency, and antenna domains. So the channel matrix 
$\bm{H}$ considered herein includes three dimensions: time, space, and antenna, and can be expressed as follows:
\begin{align}
\bm{H}[i, j, :] = h(i \Delta t, f_1+(j-1)\Delta f), 
\end{align}
where $\Delta t$, $\Delta f$, and $f_1$ denote the time interval, frequency interval, and the lowest frequency point in the frequency domain, respectively.

For the channel estimation task, a comb-type pilot pattern is used, featuring continuous pilots in the time-domain resource blocks (RBs) and a discrete arrangement in the frequency-domain RBs. The network is tasked with learning the channel characteristics across various frequency points and interpolating the missing channels at the absent frequency points. Consequently, $\bm{X}^{I}_{CE}$ and $\bm{X}^{L}_{CE}$  are defined as follows:
\begin{subequations}
\begin{align}
&\bm{X}^{I}_{CE} =  \bm{H}[1:\tilde{T}, 1:\tilde{K}/n_{pilot}:\tilde{K}, 1:\tilde{N}_t]   \label{Di1} \\
&\bm{X}^{L}_{CE} =  \bm{H}[1:\tilde{T}, 1:\tilde{K}, 1:\tilde{N}_t ]. \label{Do1}
\end{align}
\end{subequations}
Here, $\tilde{T}$ denotes the total number of timestamps, and $\tilde{K}$ denotes the number of OFDM symbols. $n_{pilot}$ denotes the number of pilots, which is typically set such that $\tilde{K}/n_{pilot}=4$. 
 
For the channel prediction task, we considered two scenarios: time-domain prediction and frequency-domain prediction, both of which help reduce the overhead of pilots. $\bm{X}^{I}_{CP}$ and $\bm{X}^{L}_{CP}$ for the time-domain prediction task are defined as:
\begin{subequations}
\begin{align}
&\bm{X}^{I}_{CP} =  \bm{H}[1:\tilde{T}, 1:\tilde{K}, 1:\tilde{N}_t]   \label{Di2} \\
&\bm{X}^{L}_{CP} =  \bm{H}[\tilde{T} + 1:\tilde{T} + \tilde{P}, 1:\tilde{K}, 1:\tilde{N}_t ],  \label{Do2}    
\end{align}
\end{subequations}
where $\tilde{P}$ represents the length of future moments to be predicted. Similarly, for the frequency-domain prediction task, $\bm{X}^{I}_{PF}$ and $\bm{X}^{L}_{PF}$ are defined as:
\begin{subequations}
\begin{align} 
&\bm{X}^{I}_{PF} =  \bm{H}[1:\tilde{T}, 1:\tilde{K} / 2, 1:\tilde{N}_t]   \label{Di3} \\
&\bm{X}^{L}_{PF} =  \bm{H}[1:\tilde{T}, \tilde{K} / 2:\tilde{K}, 1:\tilde{N}_t ]. \label{Do3}
\end{align}
\end{subequations}

\begin{figure*}[t]
    \centering
    \includegraphics[width=0.95\linewidth]{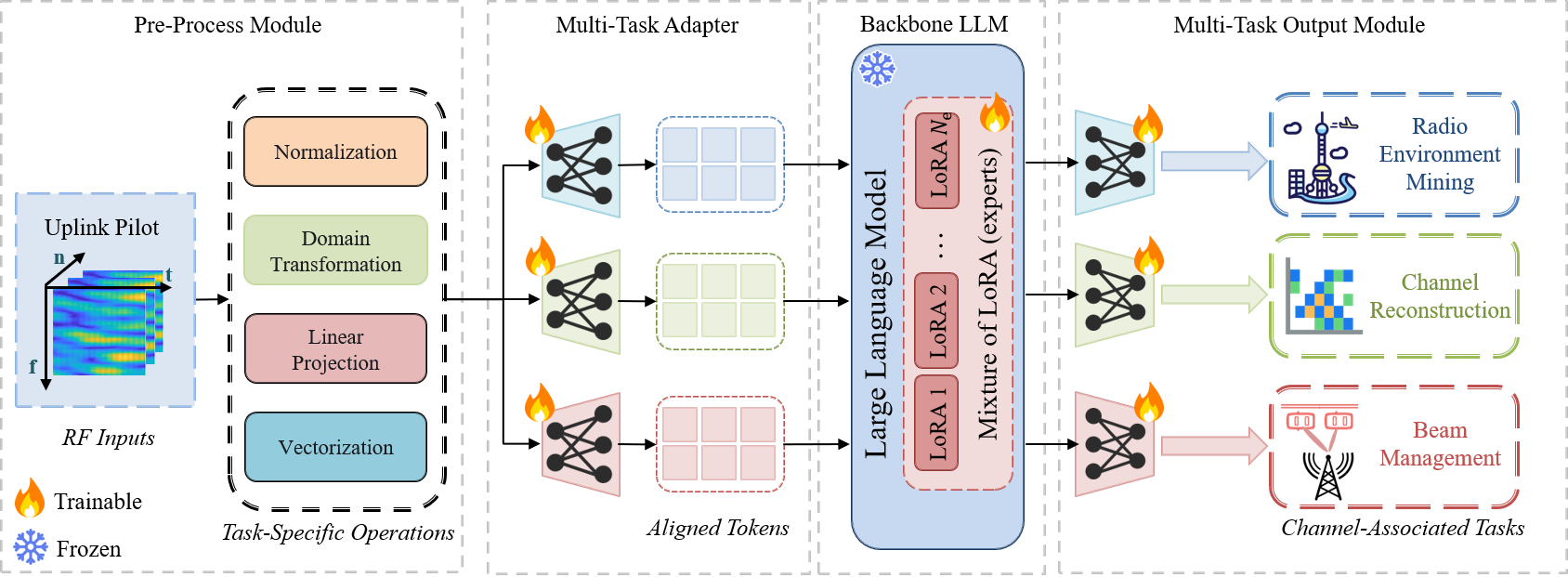}
    \caption{The proposed LLM4WM is composed of four main modules: (i) pre-process module; (ii) multi-task adapter module; (iii) backbone LLM module fine-tuned with MoE-LoRA; (iv) multi-task output module}
    \label{network}
    \end{figure*} 

\subsection{Beam Management: Sub-6G Aided Beamforming} 
Beamforming requires accurate acquisition of the optimal weight vector $\mathbf{w}_t$ based on the codebook $\mathbf{W} \in \mathbb{R}^{N_v \times N_c}$, where $N_c$ is the codebook size. To achieve higher spatial resolution, a super-resolution Discrete Fourier Transform (DFT) codebook is applied, which implies that $N_c > N_v$. Since the main paths in the sub-6G and mmWave bands are often highly correlated under Line-of-Sight (LoS) conditions, leveraging sub-6 GHz to assist mmWave beamforming is highly beneficial, while also reducing the pilot overhead for mmWave beamforming. The $I_{BF}$ and $L_{BF}$ for sub-6G aided beamforming task can be defined as:
\begin{subequations}
\begin{align}
&\bm{X}^{I}_{BF} =  \bm{H}[1, 1:\tilde{K}, 1:\tilde{N}_t]   \label{Di4} \\
&\bm{X}^{L}_{BF} =   {w}_{t}^*     . \label{Do4}
\end{align}
\end{subequations}

\subsection{Radio Environment Mining: Distance Estimation and Path Loss Estimation} 
Due to the abundant environmental information present in the acquired channel data, such as the multipath components reflecting the density of surrounding buildings, and the delay indicating the distance to the target, the task of radio environment mining aims to leverage the estimated channel to extract environmental information, such as the distance $x_d$ from the UE to the BS and the path loss $x_{pl}$ of the main path. 
This extracted information can be utilized to adjust the configuration of the communication system, ultimately improving communication quality. Thus, $\bm{X}^{I}_{DE}$ and $\bm{X}^{L}_{DE}$ for distance estimation task can be defined as:
\begin{subequations}
\begin{align}
&\bm{X}^{I}_{DE}  =  \bm{H}[1, 1:\tilde{K}, 1:\tilde{N}_t]   \label{Di5} \\
&\bm{X}^{L}_{DE}  =  x_d     . \label{Do5}
\end{align}
\end{subequations} 
Similarly, $\bm{X}^{I}_{PE}$ and $\bm{X}^{L}_{PE}$ for path loss estimation task can be defined as:
\begin{subequations}
\begin{align}
&\bm{X}^{I}_{PE} =  \bm{H}[1, 1:\tilde{K}, 1:\tilde{N}_t]   \label{Di6} \\
&\bm{X}^{L}_{PE} =  x_{pl}     . \label{Do6}
\end{align}
\end{subequations}

\section{LLM FOR Wireless Channel-Associated Tasks}
All of these tasks mentioned above are fundamentally rooted in the wireless channel, suggesting that wireless multi-task learning can significantly enhance the model's capacity to extract channel representations that generalize well. This, in turn, would improve the performance of each individual task. To leverage this potential, we propose a multi-task method for wireless channel applications, empowered by LLM, which we refer to as LLM4WM. This approach integrates the learning of these tasks within a unified network framework based on LLM. Below, a thorough description of the network components and the training process for LLM4WM are provided.
\subsection{Preprocessor Module}
Since the required channel characteristics for each task are different, a unified preprocessing approach is not conducive to fully utilizing the unique characteristics of each task. Therefore, a corresponding preprocessing function is designed for each task to preprocess the input data, and the preprocessing process for the task $n$ can be expressed as:
\begin{align}
\bm{X}_n^{pre} = f_{\text{pre},n}(\bm{X}^{I}_n),
\end{align}
where $\bm{X}_n^{pre}$ denotes the preprocessed data, and $f_{\text{pre},n}(\cdot)$ denotes the preprocessing operation for task $t$. Specifically, the preprocessing operation for the channel reconstruction tasks is tokenizing the CSI at each moment, that is, flattening the spatial and frequency features of the CSI as follows:
\begin{align}
\bm{X}_n^{pre} = \text{Flatten}(\bm{X}^{I}_n, -2),
\end{align}
where the \(\text{Flatten}(\bm{X}, i)\) operation denotes
the process of flattening the \(i\)-th dimension of the tensor \(\bm{X}\), along with all subsequent dimensions, into a single dimension.
For tasks such as beamforming, distance estimation, and path loss estimation that require channel angle features, the CSI data will undergo domain transformation to convert the spatial domain CSI to angle domain CSI, i.e.,
\begin{align}
\bm{X}_n^{pre} = \bm{X}^{I}_n \bm{F}_{\tilde{N}_n},
\end{align}
where \( \bm{F}_{\tilde{N}_t} \) is an \( \tilde{N}_t\)-dimensional DFT matrix.

\begin{figure}[htbp]
    \centering
    \includegraphics[width=0.9\linewidth]{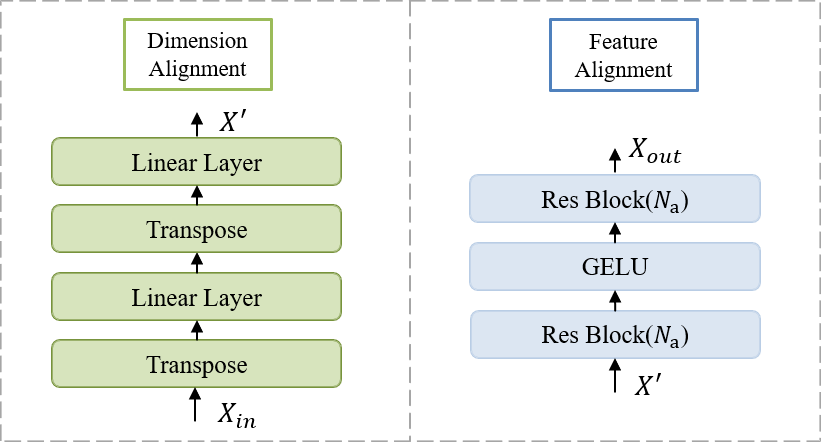}
    \caption{An illustration of the multi-task adapter module.}
    \label{adapter}
    \end{figure} 
    
\subsection{Multi-Task Adapter Module}
We extend the conventional use of adapter modules, which incorporate a small number of trainable parameters, enabling the model to preserve its existing modeling and generalization capabilities while adapting to a specific domain \cite{sung2022vl, pan2022st}. However, it is primarily designed for single-task scenarios and lacks the ability to facilitate transfer generalization across multiple tasks simultaneously.

Unlike existing single-task adapters, the multi-task adapter modules function by parallelizing multiple individual adapters to simultaneously address various tasks. The features output by each adapter are then jointly fed into LLM for multi-task learning. This design fully leverages the generalization and multi-task learning capabilities of the LLM, while the joint adaptation approach also simplifies the training process of the network. Each individual adapter within the module is assigned to a specific task, performing task alignment operations. This alignment includes both dimensional alignment and intrinsic representation alignment. As shown in Fig. \ref{adapter}, its main components include a linear alignment layer, residual feature extraction networks, and an activation function. In the individual adapter $\text{Adapter}_{n}^{in}$ for task $n$, the linear alignment layer aims to align the semantic feature space and task feature space in terms of dimensions and get the feature map as:
\begin{align}
    \bm{X}_{n}^{f} = \text{Linear}(\bm{X}_n^{pre}) \in \mathbb{R}^{L\times D_{llm}},
\end{align}
where  \(L\) and \(D_{\text{llm}}\) denote the token length of LLM's input and LLM's hidden dimension, respectively. Since the preprocessed features are two-dimensional data, the \(\text{Linear}(\cdot)\) operation includes at least two fully connected operations, which linearly map the first and second dimensions to the specified dimensions. Then the residual feature extraction networks and the activation function would act on $\bm{X}_{n}^{f}$ to obtain feature maps with aligned semantic features as:
\begin{align}
    \bm{X}_{n}^{a} = \text{Res}(\text{GELU}(\text{Res}(\bm{X}_{n}^{f}))) \in \mathbb{R}^{L\times D_{llm}},
\end{align}
where the $\text{Res}(\cdot)$ operation includes $N_{a, i}$ Res-blocks. And each Res-block contains two 1-dimensional convolution kernels and an activation function \(\text{ReLU}(\cdot)\). The convolution kernel size is $3$, and the stride is $1$. The $\text{GELU}(\cdot)$ function is a smooth, differentiable approximation of the \(\text{ReLU}(\cdot)\) function \cite{hendrycks2016gaussian}. The above steps can be simplified as:
\begin{align}
    \bm{X}_{n}^{a} = \text{Adapter}_{n}^{in}(\bm{X}_n^{pre}).
\end{align}

\subsection{Mixture-of-LoRA Based Fine-tuning}
\begin{figure}[t]
    \centering
    \includegraphics[width=1\linewidth]{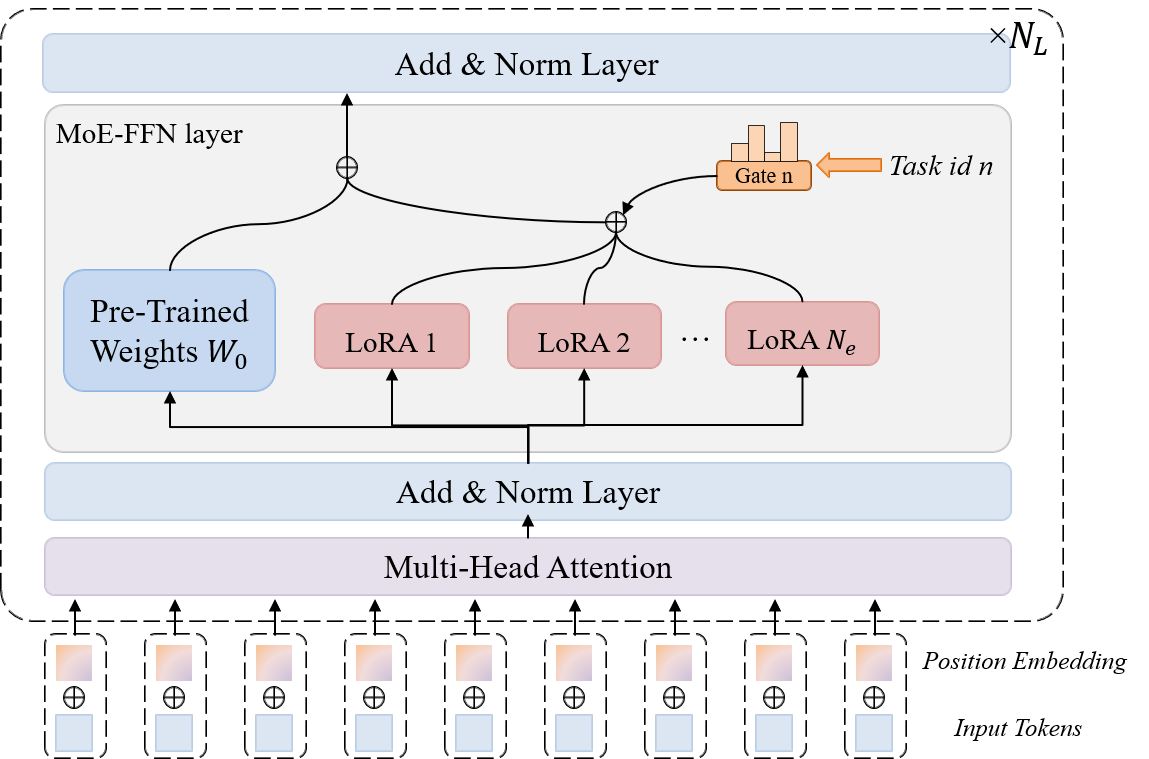}
    \caption{An illustration of the MoE-LoRA fine-tuning method.}
    \label{moe}
    \end{figure}

The backbone LLM module is essential for processing the representations extracted by the adapters. To improve the pre-trained LLM's performance on wireless channel tasks, we efficiently fine-tune its parameters using MoE-LoRA, as shown in Fig. \ref{moe}. This fine-tuning combines LoRA and MoE principles to enhance efficiency by selectively activating subsets of parameters. First, we outline the standard LoRA fine-tuning process, which trains two low-rank matrices in the model's feed-forward network and is tailored for single-task fine-tuning.

Assume the pre-trained weights are \(W_0\in \mathbb{R}^{d_{\text{out}} \times d_{\text{in}}} \), where \( d_{\text{in}} \) is the input dimension and \( d_{\text{out}} \) is the output dimension. \( A \in \mathbb{R}^{r \times d_{\text{in}}} \) and \( B \in \mathbb{R}^{d_{\text{out}} \times r} \) are two trainable low-rank matrices. Then, we can obtain the fine-tune weights \(W\in \mathbb{R}^{d_{\text{out}} \times d_{\text{in}}} \) as follows:
\begin{align}
W &= W_0 + \frac{\alpha}{r} BA,
\end{align}
where \( r \) denotes the rank of the low-rank approximation. The hyperparameter \( \alpha \) facilitates the adjustment of the rank \( r \) and is typically set as \( \alpha = 2 \times r \).

Assuming the input to the feed-forward network is \( x_t \) and the output is \( y_t \), the forward propagation of the model can be expressed as follows:
\begin{align}
y_t &= W x_t = W_0 x_t + \frac{\alpha}{r} B A x_t.
\end{align}
To extend this concept to multi-task learning, we incorporate the well-known MoE models \cite{shazeer2017outrageously}. This approach establishes a collection of independent low-rank matrices that learn task-specific features individually. A gating network is then employed to select and combine different experts for various tasks, facilitating a specific aggregation mechanism for the experts. The underlying idea is expressed as follows:
\begin{align}
y_t = W x_t = W_0 x_t + \frac{\alpha}{r} \sum_{k=1}^{N_e} {\omega}_k B_k A_k x_t,
\end{align}
where \( B_k \in \mathbb{R}^{d_{\text{out}} \times r} \) and \( A_k \in \mathbb{R}^{r \times d_{\text{in}}} \) is the \(k\)-th pair of low-rank matrices. \(N_e\) and \({\omega}_k\) represents the number of experts and \(k\)-th experts' weights, respectively. A larger value of \(N_e\)  indicates that more experts participate in the training and inference processes of the network. This can enhance the model’s capacity for accurate representation. However, it also linearly increases the training and inference costs, meaning that the specific value of  \(N_e\) requires a trade-off between model accuracy and inference speed. 

Notably, 
the design of the gating network has a direct impact on the performance of the MoE model. To prevent overfitting, we employ a single-layer linear network to generate the expert weights for each task and normalize the weight matrix using the \(\text{Softmax}(\cdot)\) function to maintain the stability of the output data. 

We apply MoE-LoRA to the linear layers within the feedforward network (FFN) of LLM while leaving the remaining parameters frozen. This approach significantly reduces the model's trainable parameters, greatly lowering training costs and improving training efficiency.

\subsection{Multi-Task Output Module}
Normal LLMs map the output features of transformer blocks to a probability distribution over the vocabulary, selecting the token with the highest probability as the output text. However, for wireless channel-associated tasks, the output is often challenging to express in text. Moreover, as the vocabulary size increases, this mapping incurs significant storage and computational costs. For instance, GPT-2's vocabulary of $50000$ words necessitates an output layer with at least $50000$ dimensions.

To address these challenges, and similar to the approach used in \cite{wu2024netllm}, we have designed a specific output layer tailored for wireless channel-associated tasks. This specialized output layer is intended to more effectively capture the target output relevant to these tasks, thereby improving performance and reducing the resource demands typically associated with large vocabulary sizes.

To align the task's output feature vector with the semantic space of the LLM, we use a multi-task adapter connected directly to the LLM's output, which is identical to the input adapter. Assuming the task $n$'s output feature of the large model is \(\bm{X}_n^{LLM}\), and the multi-task adapter for the task $n$ is denoted as \(\text{Adapter}_{n}^{out}\), so this process can be represented as:
\begin{align}
\bm{X}_n^{p} = \text{Adapter}_{n}^{out}(\bm{X}_n^{LLM})
\end{align}
where \(\bm{X}_n^{p}\) denotes output of the multi-task adapter for task $n$.

Considering that channel estimation and channel prediction tasks are more sensitive to the learning of local features, we use CNNs for processing and dimensional alignment in subsequent steps. On the other hand, tasks such as beamforming, distance estimation, and path loss estimation require obtaining a global feature representation of the channel. Therefore, the feature map will be flattened, and an MLP network will be employed for feature processing and dimensional alignment. The operations can be described as follows:
\begin{align}
\bm{X}^\text{o}_n=\left\{
\begin{aligned}
\text{CNN}(\bm{X}_n^{p}), && {n \in \{CE, CP, PF\}}\\
\text{MLP}(\bm{X}_n^{p}), && {n \in \{BF, DE, PE\}}\\
\end{aligned}
\right.
\end{align}
where \(\bm{X}^\text{o}_n\) represents prediction or estimation result of task $n$.
\subsection{Training Configuration}
 The proposed network is trained on a multi-task mixed dataset using a two-stage training approach. In the first stage, only the multi-task adapters and output layer are trained, while the LLM parameters are frozen. At this point, the model learns the mapping between the task feature space and the pre-trained LLM's text feature space. In the second stage, the LLM is fine-tuned by MoE-LoRA, while the multi-task adapters become frozen, but the output layer is still trainable. At this stage, the model leverages the LLM for joint modeling of multiple tasks, achieving better results by utilizing generalized representations across tasks. The same loss function is used for both stages, as follows:
\begin{align}
\text{Loss} = \sum_n \omega_n f_{loss, n} (\bm{X}^\text{o}_n, \bm{X}^\text{l}_n)
\end{align}

where \( f_{loss, n} \) denotes the loss function for task \( n \). And they are linearly combined with task weightings \( \omega_n \). To ensure that all tasks are well trained, we use the Dynamic Weight Average (DWA) algorithm \cite{liu2019end} to dynamically adjust each task's weight based on its loss every epoch. The selection of loss function \( f_{loss, n} \) fully considers the characteristics of the task itself. For classification problems, such as ${BF}$, we employ the cross-entropy loss function, while for regression problems, such as ${CP}$, the normalized mean square error (NMSE) \cite{jiang2022accurate} is used as the loss function.

\section{Experiments}
In this section, we first describe the simulation settings and then assess the performance of the proposed LLM4WM method from multiple perspectives: overall performance, generalization ability, and stability. We also conduct comprehensive ablation experiments to highlight the contribution of each module within the framework.

\subsection{Simulation Setup}

\subsubsection{Datasets}
We adopt the widely used channel generator QuaDRiGa \cite{jaeckel2014quadriga} to simulate time-varying CSI datasets compliant with 3GPP standards.
We consider a dual-frequency wireless system having a $1.9$ GHz sub-6G link and a $28$ GHz mmWave link. 
The hyper-parameters of the dataset generation are presented in Tab. \ref{settings}.
\begin{table}[h]
\renewcommand{\arraystretch}{1.3}
\scriptsize
\caption{Hyper-parameters for dataset generation}
\label{settings}
\centering
\begin{tabular}{|c|c|c|}
\hline
\makebox[0.06\textwidth][c]{\textbf{Parameter}}  &
\makebox[0.12\textwidth][c]{\textbf{mmWave}}  &
\makebox[0.12\textwidth][c]{\textbf{sub-6G}}  \\ \hline
Scenario  &  \multicolumn{2}{c|}{3GPP\_38.901\_UMa\_LOS} \\ \hline
Active BSs & 1 & 1 \\ \hline
Codebook size & 256 & N/A \\ \hline
Transmit antennas & 64 & 8 \\ \hline
Center frequency (GHz) & 28 & 1.9 \\ \hline
Bandwidth (GHz) & 0.5 & 0.06 \\ \hline
Antenna spacing & 0.5 & 0.5  \\ \hline
OFDM sub-carriers & 64 & 64 \\ \hline
Clusters & N/A  & 21  \\ \hline
Paths per cluster & N/A  & 20  \\ \hline
 
\end{tabular}
 
\end{table}
 
The sub-6G link operates in FDD mode to enhance spectrum utilization. Assuming that the uplink and downlink channels are adjacent, and for the uplink channel, a pilot is placed every $8$ subcarriers. For the channel prediction task, we predict future $\tilde{P}=4$ RBs based on historical $\tilde{T}=16$ RBs and set the time interval of pilots as $0.5$ ms, and for the frequency domain prediction task, the downlink channel at pilot is inferred from the uplink channel estimated or predicted by the uplink pilot.

In contrast, the mmWave link employs the TDD mode. For the sub-6G assisted mmWave beamforming task, the downlink analog precoding is also derived based on the spatial correlation of the uplink sub-6G channel estimated by the uplink pilot.
The initial position of the user is randomized and the motion trajectory is set as linear type at a speed of $30$ km/h.
The dataset contains a total of $20000$ samples. Specifically, the training set includes $15000$ samples, the validation set has $1600$ samples, and the test set consists of $3400$ samples.

\subsubsection{Baselines}
To validate the superiority of the proposed method, several model-based and deep learning-based methods are implemented as baselines.

\textbf{Traditional Methods (without deep learning)}: this class of methods does not rely on a training process but instead leverages the inherent characteristics of the channel to address specific problems. 
\begin{itemize}
 
\item \textbf{BI}: In this method, CSI is treated as a time series and bilinear interpolation (BI) is used to complete the channel reconstruction tasks.  

\item \textbf{Codebook} \cite{ali2017millimeter}: Based on spatial correlation, a super-resolution codebook is used for beam scanning in the sub-6G band to obtain the optimal mmWave downlink beam vector. This method is used to process the beam management task.

\item \textbf{FIFS} \cite{xiao2012fifs}: FIFS is a CSI-based fingerprinting system that introduces a coherence bandwidth-enhanced probability algorithm, utilizing a correlation filter to map objects to fingerprints. It is implemented for radio environment mining tasks.

\end{itemize}

\textbf{Single-task Small Model Methods}: This class of methods employs specially designed model components to address specific downstream tasks which often have a relatively small number of parameters. 
\begin{itemize}
\item \textbf{MLP} \cite{alrabeiah2020deep, ferrand2020dnn}: Many works have employed Multi-Layer Perceptron (MLP) to model complex mapping relationships in communication problems. We implement an MLP for the task of radio environment sensing and beam management.
\item \textbf{LSTM} \cite{jiang2020deep}: LSTM is designed with memory cells and multiplicative gates to deal with long-term dependency.
We implement it using $4$ LSTM layers for processing channel reconstruction tasks.
\item \textbf{CNN} \cite{safari2019deep}: A CNN-based predictor for FDD systems is proposed in \cite{safari2019deep}, where the prediction of time-frequency CSI data is treated as a two-dimensional image processing task. It contains ten convolutional layers, where the convolution kernel size is $3\times 3$.
We implement it for processing channel reconstruction tasks.

\item \textbf{WiT} \cite{salihu2022attention}: A transformer-based location estimation method, which leverages the attention mechanism to achieve robust learning effects. We implement it as described in \cite{salihu2022attention} for processing radio environment sensing tasks.

\item \textbf{Transformer} \cite{jiang2022accurate}: A transformer-based parallel channel predictor is proposed in \cite{jiang2022accurate} for TDD systems, aiming to mitigate error propagation issues. We implement it using $3$ encoders and $2$ decoders for processing channel reconstruction tasks.
\end{itemize}

\textbf{Multi-Task Small Model Methods}: This class of methods employs techniques such as low-level sharing and cross-feature fusion to enable feature sharing across different tasks, thereby achieving the functionality of a multi-purpose model.
\begin{itemize}
\item \textbf{Cross-stitch} \cite{misra2016cross}: A convolutional multi-task learning neural network equipped with “cross-stitch” unit, which could combine the activations from multiple networks. We implement it by using ResNet \cite{he2016deep} as the backbone layer. To illustrate the impact of wireless multi-task learning for small model, \textbf{Cross-stitch(s)} is added as a baseline. We implement it by directly applying the cross-stitch network while only performing a single task.
\end{itemize}

\textbf{Single-task Large Model Methods}: This class of methods typically fine-tunes large models for a single downstream task, achieving strong performance by leveraging the powerful modeling capabilities of large models.
\begin{itemize}
\item  \textbf{LLM4CP} \cite{liu2024llm4cp}: This method is the first to apply the large language model to channel prediction task through fine-tuning. We implement it and choose gpt2 as the backbone LLM and LN Tuning \cite{qi2022parameter} as the fine-tuning method for processing Channel Reconstruction tasks.
\item  \textbf{LLM4WM(s)}: A single-task fine-tuning based large model network, where we directly apply our proposed LLM4WM whilst only
performing a single task.
\end{itemize}

\subsubsection{Network and Training Parameters}
In the simulation, we adopt the Multi-Task Adapter module with $N_{a}=8$  for input and output feature alignment.
For the configuration of the MoE-LoRA fine-tuning method, we choose the number of experts to be $8$, and we set $r=8$ for each LoRA matrix.
For the output module, as mentioned above, for the specific task, we use either a three-layer MLP with 768-dimensional features or a three-layer CNN with $3x3$ kernels for the feature process. and we employ only one single-layer fully connected network to align the output dimensions.
The smallest version \cite{vaswani2017attention} of GPT-2 with $F=768$ feature dimension is adopted, the first $N_L=6$ layers of which are deployed.
Both the warm-up and cosine annealing scheduler are employed to train LLM4WM. The first $50$ epochs serve as the warm-up phase, where the learning rate increases linearly from the minimum value of $1 \times {10}^{-5}$ to $1 \times {10}^{-3}$. During the subsequent training phases, the learning rate is dynamically adjusted using the cosine annealing scheduler. Additional hyperparameters for model training are presented in Tab. \ref{networksettings}.  
\begin{table}[h]
\renewcommand\arraystretch{1.3}  
\caption{Hyper-parameters for network training}
\label{networksettings}
\centering
\scriptsize
\begin{tabular}{|c|c|}
\hline
\makebox[0.18\textwidth][c]{\textbf{Parameter}} & \makebox[0.18\textwidth][c]{\textbf{Value}} \\ \hline
Batch size  & 512 \\ \hline
Epochs & 250 \\ \hline
Optimizer & Adam (betas=(0.9, 0.999)) \\ \hline
Learning rate scheduler  & Cosine Annealing  \\ \hline
Cosine annealing period  & 100 epochs \\ \hline
Learning rate range  & \text{[$1 \times {10}^{-5}$, $1 \times {10}^{-3}$]} \\ \hline
\end{tabular}
\end{table}

\subsubsection{Performance Metric}

To evaluate performance, we employ task-specific metrics. For channel reconstruction tasks, we measure the NMSE between predictions and ground truth. The beam management task utilizes the Top-1 accuracy which shows the frequency of whether the model correctly predicts the beam index, and distance estimation and path loss estimation rely on mean absolute error (MAE) and NMSE, respectively. Meanwhile, we calculated the average metric ${\rm Avg.}$ across all tasks for each model to facilitate intuitive comparisons. The calculation of this average metric is as follows:
\begin{equation}
\resizebox{0.89\hsize}{!}{$
\begin{aligned} 
{\rm Avg.}  = \frac{1}{6} * & [ {\rm NMSE\left({CE}\right) + NMSE\left({CP}\right) + NMSE\left({PF}\right)}+ \\
      & {\rm \left(1 - Acc\left({BF}\right)\right) + MAE\left({DE}\right) + NMSE\left({PE}\right)}].
\end{aligned}
$}
\end{equation}

To provide a comprehensive evaluation of the proposed scheme's performance, we introduce an additional metric, SE, which reflects the overall performance of the communication system. SE is a crucial metric that indicates the system's achievable rate, thereby capturing the effectiveness of the communication. It is calculated by Eq. (\ref{SE}), where $\bm{h}_{t, k}$ is the actual CSI and $\bm{w}_t$ is obtained as Eq. (\ref{mf}) with predicted $\bm{h}_{t, k}$.
The communication SNR is defined as $1/{\sigma_n^2}$ and set as $10$ dB.

\subsection{Performance Evaluation}
\subsubsection{Overall Performance}
 
\begin{table*}
	\centering
        \scriptsize
        \renewcommand\arraystretch{1.5}  
	\caption{Performance of LLM4WM and other baselines: For mmWave link, the maximum SE is $9.32$ bit$\cdot$$(\text{s}\cdot\text{Hz})^{-1}$; For sub-6G link, the maximum SE is $6.33$ bit$\cdot$$(\text{s}\cdot\text{Hz})^{-1}$. The \textbf{boldface} denotes the highest score, while the \underline{underline} marks the second-best result.}
	\label{overall performance}
         
	\begin{tabular*}{1.02\textwidth}{@{\extracolsep{\fill}}c|cc|cc|cc|c|cc|c|cc|c}
		\hline
		\multirow{2}{*}{Method}&
		\multicolumn{2}{c|}{${\rm {CE}}$}&\multicolumn{2}{c|}{${\rm {CP}}$} &\multicolumn{2}{c|}{${\rm {PF}}$}& 
        \multirow{2}{*}{Method} &\multicolumn{2}{c|}{${\rm {BF}}$} &
        \multirow{2}{*}{Method} & ${\rm {DE}}$ & ${\rm {PE}}$  & \multirow{2}{*}{Avg. $\downarrow$} \cr  \cline{2-7} \cline{9-10} \cline{12-13}
        
		\multicolumn{1}{c|}{} & NMSE $\downarrow$ & \multicolumn{1}{c|}{SE $\uparrow$}   & NMSE $\downarrow$ & \multicolumn{1}{c|}{SE $\uparrow$}   & NMSE $\downarrow$ & \multicolumn{1}{c|}{SE $\uparrow$}  &\multicolumn{1}{c|}{} & Acc $\uparrow$ & \multicolumn{1}{c|}{SE $\uparrow$}  & \multicolumn{1}{c|}{} & MAE $\downarrow$ & \multicolumn{1}{c|}{NMSE $\downarrow$}  &  \cr  \hline

         {BI} & 0.654 & 5.612 & 1.796 & 2.965 & 1.293 & 5.321 &  {Codebook} & 0.288 & 7.868 & {FIFS} & 0.249 & 0.204 & 0.818 \cr
         \hline
		CNN &{0.119}&6.043&0.125&6.038&0.283&5.888&CNN&0.356&6.852&WiT&0.160&0.053&0.230 \cr
    
		LSTM &1.000&4.182&0.161&5.994&0.280&5.902&MLP&{0.831}&{8.522}&MLP&0.218&0.091&0.320 \cr 
    
        Cross-stitch(s) &0.153&5.999&0.112&6.058&0.226&5.947&Cross-stitch(s)&\underline{0.884}&\underline{8.545}&Cross-stitch(s)&0.177&0.054& 0.140 \cr 
    
		Cross-stitch&0.157&5.996&0.112&6.059&0.232&5.947&Cross-Stitch&0.858&8.525&Cross-stitch&\underline{0.131}&\underline{0.032}& 0.134\cr 
        \hline
        
        LLM4CP &\underline{0.106}&\underline{6.062}&\underline{0.106}&\underline{6.066}&0.151&6.027&LLM4CP&0.682&8.430&LLM4CP&0.199&0.122& 0.167 \cr 
    
        LLM4WM(s) &{0.108}&{6.060}&{0.106}&{6.057}&\underline{0.114}&\underline{6.061}&LLM4WM(s)&{0.878}&{8.530}&LLM4WM(s)&{0.153}&{0.052}&\underline{0.109}\cr  
        \hline
        
		\textbf{LLM4WM} &{\color{black}{\textbf{0.103}}}&{\color{black}{\textbf{6.069}}}&{\color{black}{\textbf{0.106}}}&{\color{black}{\textbf{6.068}}}&{\color{black}{\textbf{0.100}}}&{\color{black}{\textbf{6.081}}}&\textbf{LLM4WM}&{\color{black}{\textbf{0.904}}}&{\color{black}{\textbf{8.557}}}&\textbf{LLM4WM}&{\color{black}{\textbf{0.087}}}&{\color{black}{\textbf{0.028}}}& {\color{black}{\textbf{0.087}}} \cr 
        \hline
	\end{tabular*}
 
\end{table*}

Tab. \ref{overall performance} shows that LLM4WM outperforms non-learning methods, small models, and single-task fine-tuning across various tasks. Its success stems from leveraging the general knowledge of pre-trained large language models and its multi-task learning capability, which enhances feature representation. In contrast, single-task fine-tuning is prone to overfitting, leading to performance degradation in challenging scenarios. LLM4WM's combination of a Multi-Task Adapter and MoE-LoRA aligns feature spaces across tasks, resulting in more generalized performance. Our analysis of multi-task learning on large models (LM) and small models (SM) reveals that the small model suffers an average improvement of only $0.19$ dB when switching from single-task to multi-task learning, while the large model shows an improvement of $0.99$ dB. This is due to the large model's ability to extract joint representations, unlike the small model, which struggles with conflicting task knowledge. Thus, large models are better suited for handling multiple downstream tasks.

\begin{figure}[htbp]
    \centering
    \includegraphics[width=0.65\linewidth]{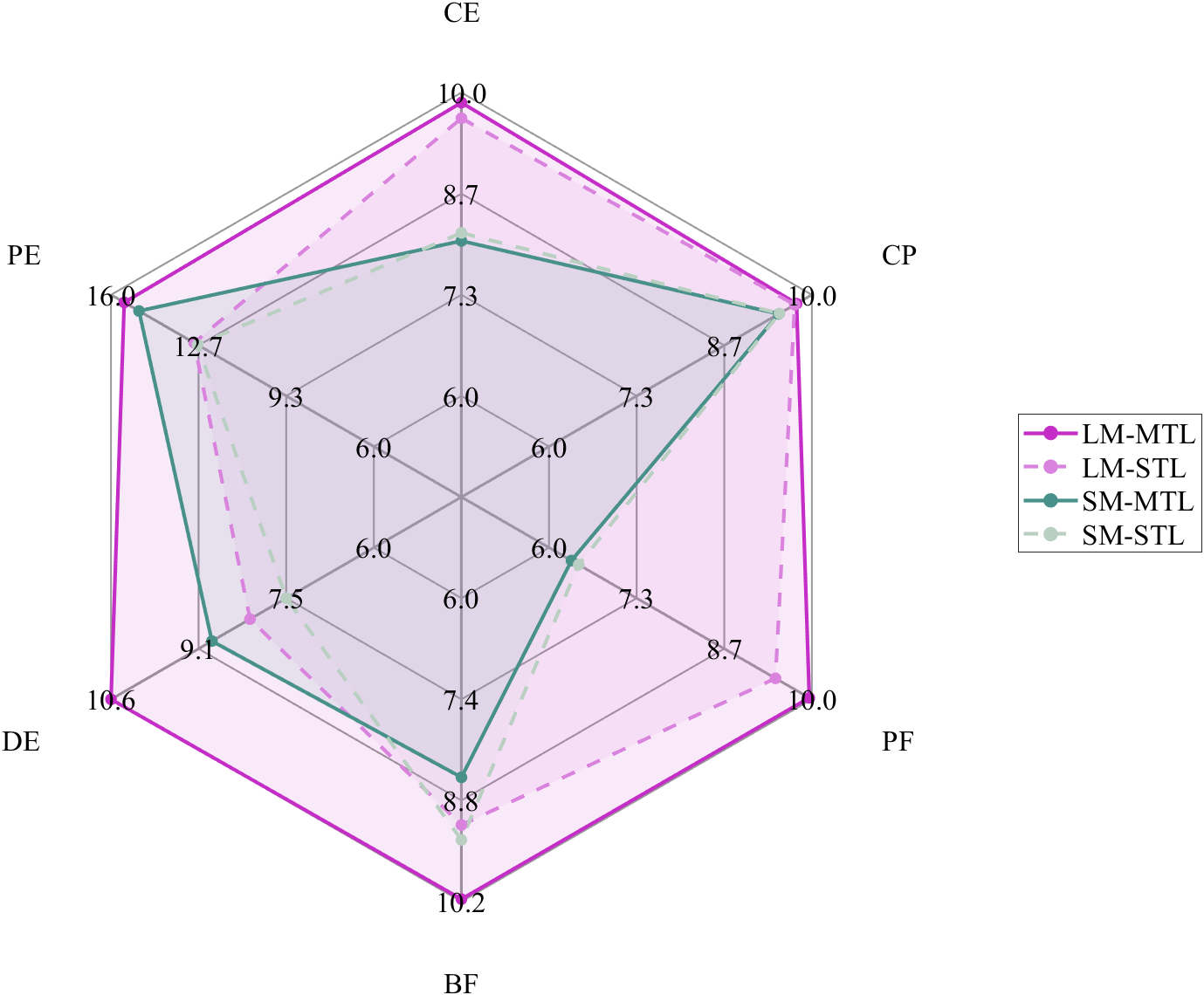}
    \caption{Performance comparison of large and small models before and after wireless multi-task learning.}
    \label{resulT_{CP}}
    \end{figure}

 \begin{figure}[htbp]
    \centering
    \includegraphics[width=1\linewidth]{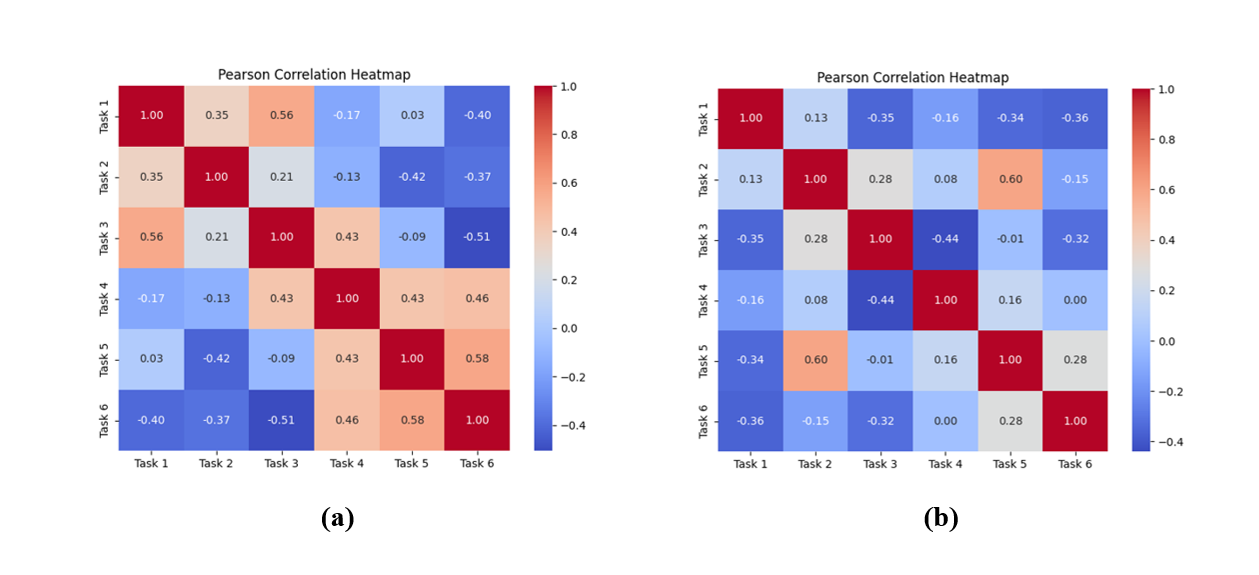}
    \caption{Pearson correlation coefficient heatmap of expert combination weights for various tasks}
    \label{result_Pearson_correlation}
    \end{figure}

To verify whether the experts in the MoE are effectively allocated based on task types, we use the Pearson correlation coefficient as a metric. Heatmaps of expert combinations weights from two randomly selected MoE-LoRA layers are shown in Fig. \ref{result_Pearson_correlation}. The results reveal that the correlation between expert combination weights for most tasks is quite low, indicating that the gating network indeed learns distinct expert combinations for different task types. Additionally, it can be observed that tasks with similar characteristics exhibit higher correlations, likely due to the fact that neighboring tasks often belong to the same task set, thus exhibiting stronger correlations.

\subsubsection{Generalization Experiments}

\begin{table*}
	\centering
        \scriptsize
        \renewcommand\arraystretch{1.5}   
	\caption{Generalization Performance of LLM4WM and Other Baselines: the average loss across tasks is computed and presented in the final column labeled "AVG.". The \textbf{boldface} denotes the highest score, while the \underline{underline} marks the second-best result.}
	\label{generalization performance}
	\begin{tabular*}{0.87\textwidth}{@{\extracolsep{\fill}}c|c|c|c|c|c|c|c|c|c|c|c}
		\hline
		\multirow{2}{*}{Train Set}&\multirow{2}{*}{Test Set}& 		\multirow{2}{*}{Method}&
		\multicolumn{1}{c|}{${\rm {CE}}$}&\multicolumn{1}{c|}{${\rm {CP}}$} &\multicolumn{1}{c|}{${\rm {PF}}$}& 
        \multirow{2}{*}{Method} &\multicolumn{1}{c|}{${\rm {BF}}$} &
        \multirow{2}{*}{Method} & ${\rm {DE}}$ & ${\rm {PE}}$  & \multirow{2}{*}{Avg. $\downarrow$}  \cr  \cline{4-6}  \cline{8-8} \cline{10-11}
		&  &  & NMSE $\downarrow$ &  NMSE $\downarrow$ &  NMSE $\downarrow$ &  & Acc $\uparrow$ & & MAE $\downarrow$ & NMSE $\downarrow$ &  \cr
		\cline{1-12}
        
		\multirow{8}{*}{ \makecell[c]{UMa \\ 1.9GHz }} & \multirow{4}{*}{ \makecell[c]{RMa \\ 1.9GHz }}
		& \textbf{LLM4WM} &\textbf{0.143}&\underline{0.145}&\textbf{0.162}&\textbf{LLM4WM}&\textbf{0.413}&\textbf{LLM4WM}&\textbf{0.336}&\underline{0.285}&\textbf{0.276} \cr
        \cline{3-12}
        & &LLM4CP&\underline{0.177}&\textbf{0.133}&\underline{0.292}&LLM4CP&0.306&LLM4CP&0.370&0.311&\underline{0.330} \cr
        \cline{3-12}
        & & CNN &0.187&0.137&0.384&CNN&0.215&WiT&\underline{0.339}&\textbf{0.220}&0.376 \cr
        \cline{3-12}
        & & LSTM &1.000&0.309&0.545&MLP&\underline{0.365}&MLP&0.539&0.473&0.584 \cr
        \cline{2-12}

        & \multirow{4}{*}{ \makecell[c]{UMa \\ 2.4GHz }}
		& \textbf{LLM4WM} &\textbf{0.101}&\textbf{0.110}&\textbf{0.135}&\textbf{LLM4WM}&\textbf{0.785}&\textbf{LLM4WM}&\textbf{0.126}&\textbf{0.047}&\textbf{0.122} \cr
        \cline{3-12}
        & &LLM4CP&\underline{0.110}&\underline{0.113}&\underline{0.196}&LLM4CP&0.685&LLM4CP&0.182&0.073&\underline{0.165} \cr
        \cline{3-12}
        & & CNN &0.115&0.121&0.381&CNN&0.375&WiT&\underline{0.143}&\underline{0.047}&0.239 \cr
        \cline{3-12}
        & & LSTM &1.000&0.174&{0.340}&MLP&\underline{0.769}&MLP&0.256&0.134&0.356 \cr
        \cline{1-12}
	\end{tabular*}
\end{table*}

Generalization, referring to the ability of models to maintain performance in new communication scenarios, is crucial for real-world deployment as it reduces the need for frequent updates. We use only $10\%$ of the RMa dataset to transfer a model trained in the UMa scenario, as well as a model trained on the $1.9$ GHz sub-6 GHz link dataset to the $2.4$ GHz sub-6 GHz link dataset. Results in Tab. \ref{generalization performance} indicate that, despite the challenges of multi-task generalization and transfer, our approach consistently outperforms others across most tasks. The slight dip in radio environment mining performance is due to the task's relative simplicity in the LOS scenario, where smaller models like WiT excel. However, our model excels in more complex tasks like channel estimation, which requires understanding multidimensional features, further confirming that large models are better suited for dynamic real-world communication scenarios.

\subsubsection{Hyper-parameter Analysis}

 \begin{figure}[htbp]
    \centering
    \includegraphics[width=1.0\linewidth]{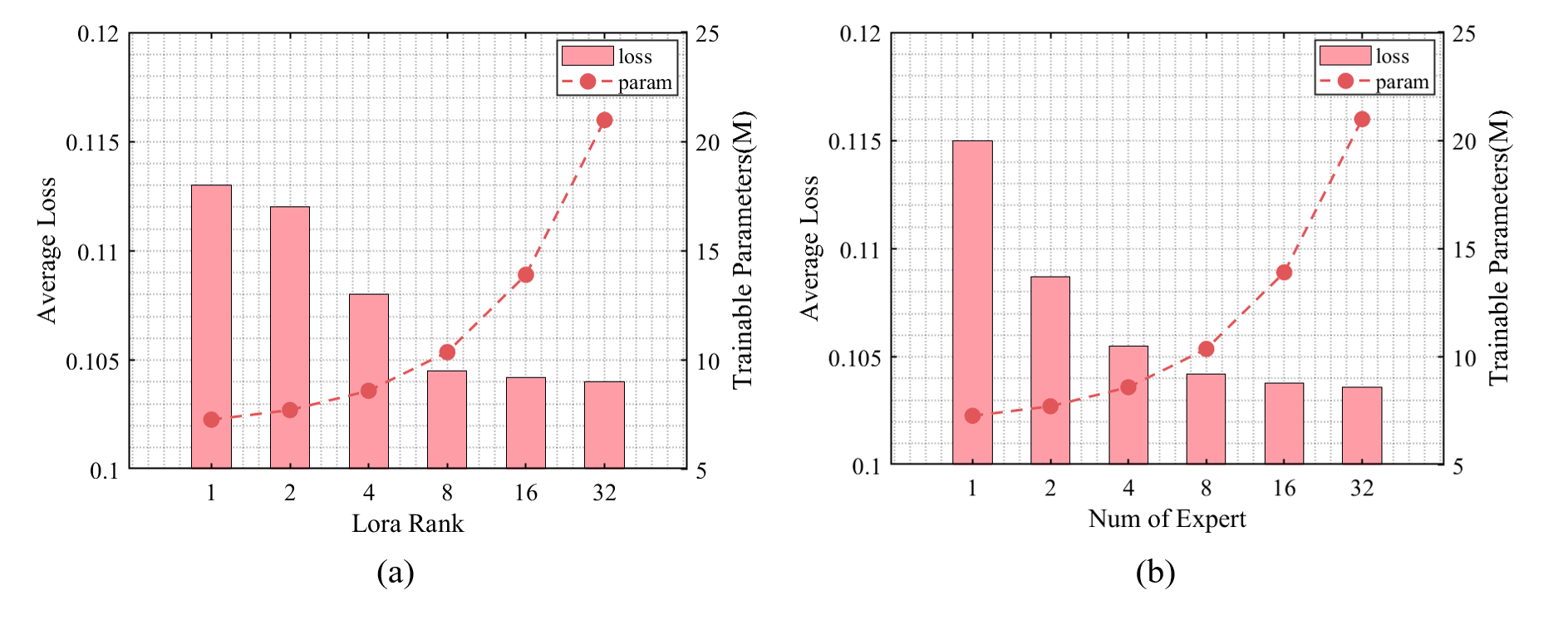}
    \caption{The performance of LLM4WM under different Lora ranks and number of experts.}
    \label{result_hyper_parm}
    \end{figure} 

To illustrate the rationale behind the hyperparameter settings, we conduct a thorough experiment on how hyperparameters affect the performance of LLM4WM. Specifically, we examine the effects of varying the LoRA rank and the number of experts, as depicted in Fig. \ref{result_hyper_parm}. When we increase the LoRA rank while keeping the number of experts fixed at $8$, the performance of LLM4WM gradually improves. This can be attributed to the enhanced adaptability of the model to the data distribution due to the increase in trainable parameters. However, this improvement comes with higher training overhead as a cost. After weighing the trade-off between performance and computational efficiency, we determine that a LoRA rank of $8$ provided the optimal balance. Subsequently, with the LoRA rank fixed at its optimal value of $8$, we incrementally increase the number of experts. A similar trend to that of the LoRA rank was observed, as an increase in the number of experts effectively enhanced the model's analytical and representational capacity. Balancing performance and computational efficiency, we determine that setting the number of experts to $8$ was the most appropriate choice.
 
\subsubsection{Ablation Experiments}
     
\begin{table*}[t]
\renewcommand\arraystretch{1.5}  
\caption{Test results of ablation experiments for multi-task adapter module and the backbone LLM module.}
\label{result_ablation}
\centering
\footnotesize
\begin{tabular*}{0.9\textwidth}{c@{\extracolsep{\fill}}ccccccc}
\hline
  Metric & LLM4WM & w/o $\text{Adapter}_\text{in}$ & w/o 
 $\text{Adapter}_\text{out}$ & w/o Adapter & w/o LLM & Frozen LLM    \\ \hline
Average Loss & {0.087} & 0.092& 0.095 & 0.102 & 0.117 & 0.092 \\  
Loss Increase Ratio & {0.00\%} & 6.50\% & 9.54\% & 17.62\% & 34.40\% & 6.15\%
  \\ \hline
\end{tabular*}
 
\end{table*}

To assess the effectiveness of the proposed modules, ablation experiments are conducted by altering or removing the configurations of the multi-task adapter and backbone LLM modules. Variations for the multi-task adapter include w/o $\text{Adapter}_\text{in}$ (placing adapter only on the output side of the LLM), w/o $\text{Adapter}_\text{out}$ (placing adapter only on the input side of the LLM), and w/o Adapter (no adapters). For the backbone LLM, variations include w/o LLM (removing the large model) and frozen LLM (freezing pre-trained weights). Results in Table V show that all ablation configurations led to performance declines, highlighting the effectiveness of both the multi-task adapter and backbone LLM modules. Notably, the impact of removing the backbone LLM is significantly greater, indicating its critical role in the success of multi-task joint learning for wireless tasks.

\begin{table*}[t]
\footnotesize
\centering
\renewcommand\arraystretch{1.5}  
\caption{Network parameters (trainable parameters/total parameters) and the interference cost per batch.}
\label{cost}
\begin{tabular*}{0.9\textwidth}{c@{\extracolsep{\fill}}ccccccc}
\hline
Metric & MLP & CNN & LSTM & WiT & LLM4CP & LLM4WM  \\ \hline
Trainable Network parameters (M) &  1.29 & 2.14 & 1.17 & 19.19 & 1.80 & 1.13 \\  

Total Network parameters (M) & 1.29 & 2.14 & 1.17 & 19.19 & 82.91
 & 88.71 \\  
Interference time (ms) & 0.32 & 0.49 & 6.49 & 2.97 & 8.62 & 6.00 \\ \hline
\end{tabular*}
\end{table*}
\subsubsection{Efficiency Evaluation}
We evaluate the model's training and inference cost of LLM4WM with other baselines to assess the difficulty of deploying the model in practical scenarios, as shown in Tab. \ref{cost}.
The same machine with $4$ Intel Xeon Platinum 8375C CPUs, $4$ NVIDIA GeForce RTX4090 GPUs, and $256$ GB of RAM is used to conduct all evaluations.
To facilitate representation and comparison, the data in the table reflects the average performance across tasks for each method. 
Notably, the MoE-LoRA fine-tuning method results in LLM4WM having trainable parameters comparable to those of smaller models. This highlights both the training efficiency and parameter efficiency of LLM4WM, as adding new tasks increases the model’s parameters by only about $1.13$ M, a small fraction of the total. Additionally, utilizing a lightweight backbone LLM ensures that the overall inference speed of LLM4WM remains acceptable. Thus, LLM4WM demonstrates significant potential for deployment in future communication scenarios marked by increasing demand and the need for customized services involving numerous tasks.

\section{CONCLUSIONS}
In this paper, we have proposed a novel multi-task fine-tuning framework tailored for large models in wireless communication systems. By leveraging a diverse multi-task dataset, our approach has enabled the model to perform various wireless channel-associated tasks concurrently. To facilitate the extraction of shared representations across multiple tasks, we integrated MoE-LoRA into the fine-tuning process, empowering the backbone model to dynamically adapt by optimally combining expert modules and improving task-specific performance. Additionally, we employed a multi-task adapter to harmonize the feature spaces of different tasks with the semantic embedding space of the large model, ensuring coherent task alignment. Preliminary simulation results have demonstrated the robust multi-task learning and generalization capabilities of the proposed LLM4WM framework. Furthermore, ablation studies have underscored the critical contributions of each module to overall system performance. The expert weight heatmap has validated the efficacy of the MoE mechanism in adaptively allocating expert resources, highlighting its role in enhancing model specialization and flexibility.
  
\bibliographystyle{IEEEtran}
\bibliography{LLM4WM}

\end{document}